\documentstyle[12pt]{article}
\begin{document}
\begin{titlepage}
\title{Comment on 'Quantum Backreaction on "Classical" Variables'}
\author{Lajos Di\'osi\\
        KFKI Research Institute for Particle and Nuclear Physics\\
        H-1525 Budapest 114, POB 49, Hungary\\
        e-mail: diosi@rmki.kfki.hu\\\\
        {\it bulletin board ref.: quant-ph/9503003}}
\date{2 March 1995}
\maketitle
\begin{abstract}
It is argued that the bracket of Anderson's canonical theory should have been
antisymmetric otherwise serious controversies arise like violation of
both hermiticity and the Leibniz rule of differentiation.
\end{abstract}
\end{titlepage}

In his recent Letter [1], Anderson proposed a canonical formalism to couple
quantum and (quasi-)classical dynamic variables. Although the proposal might
really promise good physics (cf. Ref.~[2]) its mathematical realization seems
questionable. In my opinion, the author takes too easy that his
quasi-classical bracket (2) is {\it not} antisymmetric. In fact, the lack of
antisymmetry leads, in due course, to unacceptable consequences for time
evolution of  dynamic variables.

Consider the equation of motion (4) of the Letter. It will violate the
Leibniz rule of differentiation as well as hermiticity of the dynamic
variable $A$. In Anderson's first example, the Hamiltonian is ${1\over2}kp^2$
and yields $\dot q=kp$ and $\dot x={1\over2}p^2$  for the time derivatives
of the canonical coordinates. From them, applying Leibniz rule first,
we can calculate the (initial) time derivative of the dynamic variable
$A=xq+qx$ and obtain
\hbox{$\dot A=\dot xq+x\dot q +\dot qx+q\dot x
             = {1\over2}p^2q + {1\over2}qp^2 + 2xkp$}.
If we calculated $\dot A$ directly from the equation of motion (4) we would
obtain a different expression \hbox{$\dot A=qp^2+2xkp$}. It is hardly an
acceptable result since it is {\it not} hermitian and the Leibniz rule fails
obviously to hold.

Similar effects will occur quite generally. Consider, e.g., a quantum
particle and another (quasi-)classical one, interacting via translation
invariant potential $V(q-x)$. The Letter's Eq.~(4) preserves the total
momentum $p+k$ but it leads to antihermitian time derivative $-i\Delta V(q-x)$
when applied to the {\it square} $(p+k)^2$ of the total momentum. Anderson
himself notices that, e.g., the energy of a conservative system might not be
conserved in his theory.

These controversies would not arise at all had we chosen the antisymmetric
bracket of Aleksandrov [3] and of Boucher and Traschen [4]:
$$[A,B]_{q-c}=[A,B]+{i\over2}\{A,B\}-{i\over2}\{B,A\}$$
instead of the Letter's choice (2). I admit that I have failed to see enough
reason of Anderson's departures from the above bracket. Especially, since
the antisymmetric bracket can even be {\it derived} from quantum mechanics
in proper (quasi-)classical approximation as shown by Aleksandrov [3].
This should be a maximum justification in favor of the antisymmetric bracket
even if the Letter's algebraic construction happened to result in a
consistent theory.

\bigskip
This work was supported by OTKA Grant No.1822/1991.


\begin{thebibliography}{99}
\bibitem{And} A. Anderson, Phys.Rev.Lett. {\bf 74}, 621 (1995).
\bibitem{Mad} J. Maddox, Nature {\bf 373}, 469 (1995).
\bibitem{Ale} I.V. Aleksandrov, Z.Naturforsch. {\bf 36A}, 902 (1981).
\bibitem{Bou} W. Boucher and J. Traschen, Phys. Rev. D {\bf 37}, 3522 (1988).
\end{thebibliography}
\end{document}